\documentstyle[12pt]{article}

\input psfig

\def\xslide#1#2#3#4#5#6#7{\centerline{\psfig
{figure=#1,height=#2,bbllx=#3bp,bblly=#4bp,bburx=#5bp,bbury=#6bp,width=#7,clip=}}}

\title{Further study of the $\pi\pi$ \mbox{ S--wave }\\
isoscalar  amplitude below the \KK threshold}
\author{R. Kami\'nski
\thanks{temporarily at LPNHE et LPTPE, Universit\'es
D.~Diderot et P.~et~M.~Curie, 4, Place Jussieu,
75252 Paris CEDEX 05, France}
, L. Le\'sniak and K. Rybicki \\ 
\small {Henryk Niewodnicza\'nski Institute of Nuclear Physics,}
 \\ \small{ PL 31-342 Krak\'ow, 
Poland}}

\newcommand{\be}{\begin{equation}}
\newcommand{\ee}{\end{equation}}
\newcommand{\ba}{\begin{eqnarray}}
\newcommand{\ea}{\end{eqnarray}}
\newcommand{\pipi}{$\pi\pi$ }
\newcommand{\KK}{$K\overline{K}$ }
\newcommand{\fo}{$f_0(980)$ }
\newcommand{\ro}{$\rho(770)$ }
\newcommand{\fd}{$f_2(1270)$ }
\newcommand{\rf}{$\rho_3(1690)$ }

\newcommand{\aone}{$a_1$ }
\newcommand{\ep}{$f_0(750)$ }
\newcommand{\fgg}{$f_0(1500)$ }
\newcommand{\mpp}{$m_{\pi\pi}$ }

\newcommand{\et}{$\eta$ }

\newcommand{\downp}{"down--flat" }
\newcommand{\downs}{"down--steep" }
\newcommand{\upp}{"up--flat" }
\newcommand{\ups}{"up--steep" }

\begin{document}

\maketitle

\begin{abstract}
 
We continue the analysis of $S$--wave production amplitudes for the 
reaction $\pi^- p \rightarrow \pi^+ \pi^- n$ involving the data
obtained by the CERN--Cracow--Munich collaboration on a transversely 
polarized target at 17.2 GeV/c $\pi^-$ momentum. This study deals with 
the region below the $K\overline{K}$ threshold. In particular,
we study the \ups solution containing a narrow S--wave resonance
under the \ro . This solution exhibits a considerable inelasticity $\eta$ which
does not have any 
 physical interpretation. Assuming that this inelasticity behaviour represents
 an unlikely fluctuation
we impose $\eta\equiv 1$ for all data points. This leads to 
non-physical results in one third of the $\pi^+ \pi^- $ effective mass bins and 
in the remaining mass
 bins some parameters behave in a queer way. The situation is even worse for the
  \downs solution.
 We conclude that the 17.2 GeV data
 cannot be described by a relatively narrow $f_{0}(750)$. The \downp and \upp
solutions which easily pass the $\eta\equiv 1$ constraint exhibit a slow 
increase of phase shifts in the \ro mass range.

\end{abstract}
 
 
\section{Introduction \label{introd}}
 
\hspace{0.6cm}
Scalar mesons are one of the main puzzles of light quark 
spectroscopy. Even in the lowest mass region (below the $K\overline{K}$
threshold) the situation is far
from being clear. In addition to a broad $f_0(500)$ interpreted either as
a $q-\overline{q}$ object (see e.g. Ref. \cite{Torfras}) or as a 
glueball by Ochs \cite{minkochs}, the relatively narrow  $f_0(750)$ has been 
persistently claimed by Svec [3-5]
. Arguments against the
narrow $f_0(750)$ has been given e.g. by Morgan in Ref. \cite{morgan}. 
The main source of information in this mass region 
is the \pipi partial wave analysis (PWA) yielding the $S$--wave.
It should be stressed that a study of $S$--wave objects {\it does require} the 
partial wave analysis to "subtract" the dominant contribution  of leading 
\ro meson.\\ 
\hspace*{0.6cm}Virtually all PWA's in last decades were based on the old  CERN--Munich 
experiment \cite{grayer}, 
which supplied 3$\times$10$^5$ events of the reaction
\be
\pi^-p\rightarrow \pi^+\pi^- n
\label{reaction}
\ee
at 17.2 GeV/c. The number of observables provided by
such experiment is much smaller than the number of real parameters needed to 
describe the partial waves. Consequently, the dominance of pseudoscalar 
exchange, equivalent to the absence of pseudovector exchange
 and several other physical assumptions have been made
in previous studies [7-8]
. These results have been generally used
{\it without even mentioning the assumptions} essential for their 
derivation.\\ 
\hspace*{6mm}In our previous paper \cite{klr} (hereafter called 
paper I) we have used the results of PWA performed in the the effective
mass \mpp range 
from 600 MeV to 1600 MeV at four-momentum transfer squared
$|t|=(0.005-0.200)$ GeV$^2$/c$^{2}$ using additionally
the results of the  polarized target experiment. This experiment, performed 
25 years ago by  the CCM (CERN--Cracow--Munich) collaboration \cite{becker1}, 
provided 1.2$\times$10$^6$ events of the reaction
\begin{equation}
\pi^-p_{\uparrow}\rightarrow \pi^+\pi^- n
\label{reaction_pol}
\end{equation}
also at 17.2 GeV/c. Combination of results of both experiments yields 
a number of observables sufficient for performing a quasi--complete and 
energy independent PWA. This analysis is only quasi--complete  because of an 
unknown phase between  two sets of transversity amplitudes. Nevertheless, 
{\it full} (containing both $\pi$ and \aone exchange) intensities of 
partial waves could be determined in a model--independent way. 
The original study of the CCM collaboration \cite{becker2} 
removed ambiguities appearing in  earlier studies, except for the  
"up--down" ambiguity \cite{mms}. 
The "up" solution contains an $S$--wave resonance  just under the \ro and of 
similar width, while the "down" $S$--wave modulus stays high and 
nearly constant all the way to the \fo. \\ 
\hspace*{6mm}In paper I we have made a further step in the analysis of 
17.2 GeV/c data bridging two sets of transversity amplitudes. We required the 
phases of the leading $P$, $D$ and $F$--transversity amplitudes to follow 
the phases of the Breit--Wigner \ro, \fd and \rf resonant amplitudes in the low,
 medium 
and high mass region, respectively. Further, using the measured phase 
differences between the $S$--wave and the higher waves we determined the 
{\it absolute} phases of the  $S$--wave transversity amplitudes. Once the 
phases are known, the  $S$--wave amplitudes of different transversity can be 
combined which allows us to determine explicitly for the first time the 
pseudoscalar and pseudovector exchange amplitudes in the $S$--wave. This has 
been done using much weaker assumptions\footnote{the main assumption was 
neglecting of a possible influence of the \aone exchange in $I=2$ $S$--wave as 
well as in $P$, $D$ and $F$ waves.}  than those made in {\it any} earlier 
analysis. The price we pay is a fourfold ambiguity in our pseudoscalar 
exchange $S$--wave amplitude. In addition to "up-down" ambiguity of the old 
CCM analysis \cite{becker2} there are ambiguities resulting from adding or 
subtracting the phase difference since the PWA yields only the absolute value
of the phase difference.
Thus we have "down-flat", "down-steep", "up-flat" and "up-steep" solutions.
Differences between "flat" and "steep" refer mainly to the behaviour
of the $S$--wave phase shifts below the \fo . Above the \fo all the 
solutions are fairly similar. It is
the region below the \fo which is a subject of the present paper.
The main difference is that both "steep" solutions contain a relatively
 narrow\footnote{hereafter "relatively narrow" means "with a width close to 
$\Gamma_\rho = 150$ MeV".} resonance under the \ro  (like the old "up" solution)
 while both "flat"
solutions indicate a $f_0(500)$ state with a width of about $500$ MeV. 
In particular the "down-flat" solution is very similar
to the old solution of the CERN-Munich group \cite{grayer}.\\
\hspace*{6mm}In paper I we have determined inelasticities $\eta$ for 
isoscalar $\pi$-exchange amplitudes in all solutions (see Fig. 1). It is 
obvious that both "flat" solutions easily pass the $\eta\le 1$ test and
the "down-steep" solution is not acceptable. Unfortunately, the situation
is not as simple as presented in the last edition of Review of Particle 
Properties \cite{RPP}. The authors of "Note on scalar mesons" discussing 
a hypothetical narrow state at $750~$ MeV write: "Such a solution is also 
found by (KAMINSKI 97)...However they show that unitarity is violated for
this solution; therefore a narrow light $f_{0}$ state below $900~$ MeV seems
to be excluded". The point is that our \ups  solution although exhibiting
"a puzzling behaviour of inelasticity" cannot be excluded so simply as our
\downs solution and it also contains a narrow $f_{0}(750)$. It should be recalled, 
that contrary to the Svec analysis which uses only moduli of the unseparated
(pseudoscalar and pseudovector exchange) $S$--wave, we study inelasticity
and phase shift of the pure  $\pi\pi\rightarrow\pi\pi$ isoscalar 
 \mbox{$S$--wave}.
In this paper we discuss the feasibility of an interpretation of 17.2 GeV data
in terms of a narrow $f_{0}(750)$.\\
\hspace*{6mm}The paper is organized as follows. In Sect. 2 we study the 
inelasticity behaviour in more detail. In Sect. 3 we impose strict 
$\eta \equiv 1$ 
condition on all solutions for the effective $\pi\pi$ masses below the
\KK threshold. The results are discussed in Sect. 4 and summarized
 in Sect. 5.
 
\section{$S$--wave inelasticity \label{inelas}}
 
\hspace{0.6cm}As seen in Fig. 1a the inelasticity for the \ups solution
behaves in a non-trivial way. Below $720~$ MeV and above $820~$ MeV $\eta>1$ 
while in the intermediate mass region $\eta<1$.
Qualitatively this behaviour of the "up-steep" solution is very
similar to the "down-steep" solution. The latter solution was
excluded in paper I since the corresponding inelasticity significantly
exceeded unity for \mpp $>$ 820 MeV. A general trend of $\eta$ points for
 the "up-steep"
solution resembles very much a trend of the "down-steep" data
in the region below 820 MeV. Above 820 MeV the "up-steep"
values of inelasticity lie also above unity. They are, however,
closer to unity than the "down-steep" inelasticities. This fact
prevented us to reject in paper I the "up-steep" solution solely on a basis
of the minimization fit of the sum $\sum_{i}
(\eta_{i}-1)^2$ in the whole energy region between 600
MeV and 1000 MeV.

Now we examine in detail the \mpp dependence of the
inelasticity $\eta$ corresponding to the four solutions.
Let us define by $a$, $b$ and $c$~ three ranges of $m_{\pi\pi}$: $600
\le m_{\pi\pi} \le 720$ MeV, ~$ 720 \le m_{\pi\pi} \le 820 $ MeV
and $820 \le m_{\pi\pi} \le 1000 $ MeV, respectively. Then we make a simple fit
to inelasticities in the above three ranges by a constant N. We minimize the sum
$\sum_{i}
(\eta_{i}-N)^2/ \Delta {\eta_i}^2$, where $\Delta {\eta_i}$ are the 
experimental errors of $\eta$. The results are shown in Table 1.
Let us notice that for the "steep" solutions the second value in the
intermediate region $b$ is definitely lower than unity, this is also different
from the values in the ranges $a$ and $c$ which in turn are
higher than unity. In this aspect the "up-steep" solution is very
similar to the solution "down-steep" already rejected in paper I.
On the other hand both solutions "down-flat" and "up-flat" do
not exhibit any dip of $\eta$ in the range $b$; the constants N are very close 
to unity everywhere. If we determine one common value of N in the whole range 
between 600 MeV and 1000 MeV then we obtain 1.00$\pm$0.06 for the solution
"down-flat" and 0.98$\pm$0.06 for the solution "up-flat". Both are compatible 
with unity as seen in Fig. 1b. 
\begin{table}[htb]
\centering
\caption{ Constants N fitted to $\eta$ in three ranges:  
$600 < m_{\pi\pi} < 720$ MeV ($a$), 
$720 < m_{\pi\pi} < 820$ MeV  ($b$)
and $820 < m_{\pi\pi} < 1000$ MeV ($c$)}

\vspace{.7cm}

\begin{tabular}{|l|c|c|c|}
\hline
Solution & $a$ & $b$ & $c$ \\
\hline
"up-steep" & 1.17 $\pm$ 0.10 & 0.67 $\pm$ 0.17 & 1.18 $\pm$ 0.10 \\
"down-steep" & 1.40 $\pm$ 0.12 & 0.82 $\pm$ 0.10 & 1.67 $\pm$ 0.11 \\
"up-flat" & 0.97 $\pm$ 0.10 & 1.02 $\pm$ 0.11 & 0.95 $\pm$ 0.12 \\
"down-flat" & 0.97 $\pm$ 0.13 & 1.05 $\pm$ 0.09 & 0.96 $\pm$ 0.09 \\
\hline
\end{tabular}
\label{abc}
\end{table}
 

\begin{figure}
    \begin{center}
\xslide{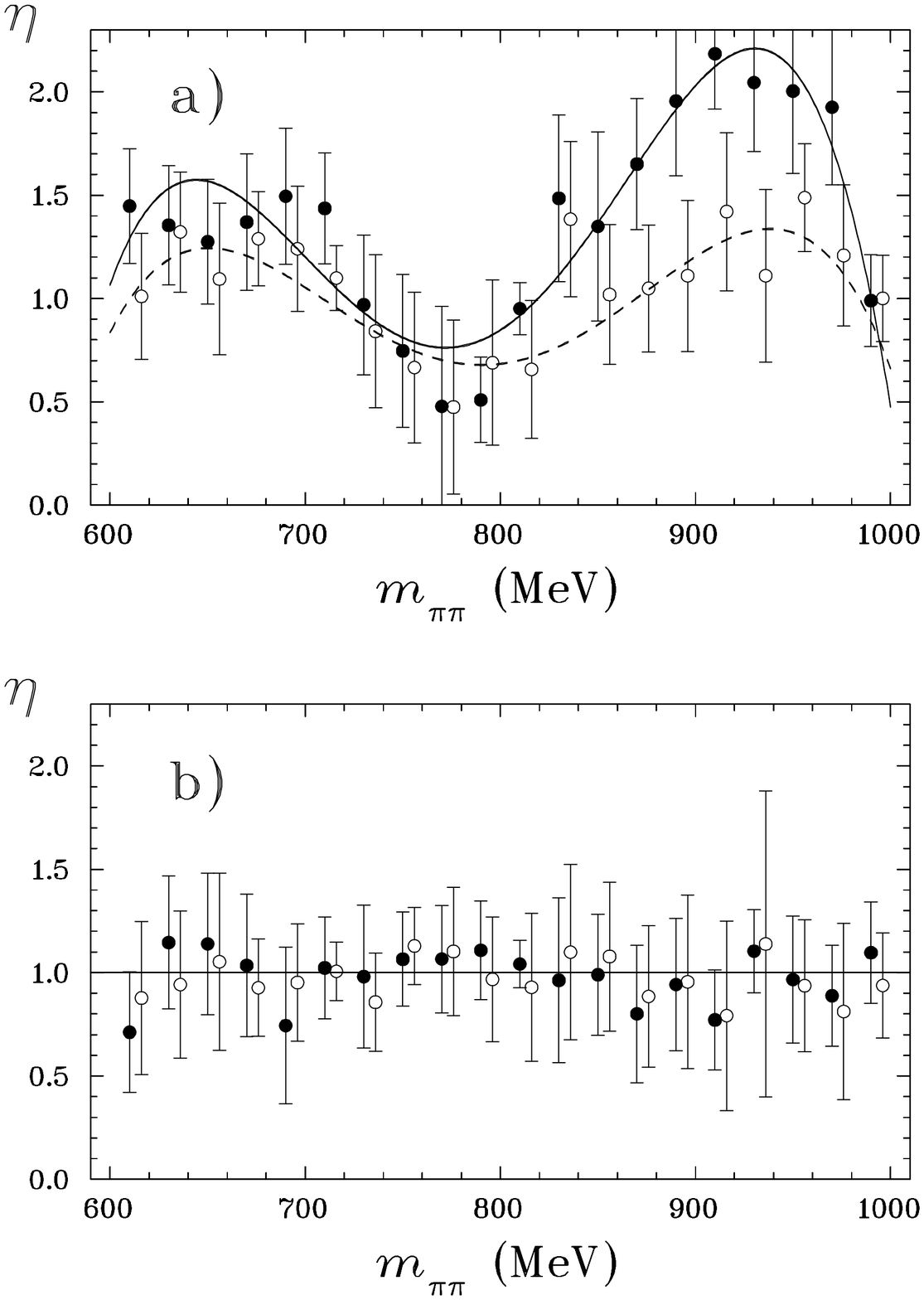}{17cm}{70}{50}{555}{732}{14cm}

\vspace{0.6cm}

\caption{{\bf a)} Scalar--isoscalar \pipi inelasticity coefficient
\et versus the effective \pipi mass for the \downs (full circles) and 
\ups (open circles, shifted by 6 MeV) solutions.
 Solid and dashed lines represent fits of the
 fourth order polynomial to the coefficient \et for the \downs and \ups 
 solutions, respectively. {\bf b)} Same as in a) but for the \downp (full 
 circles) and \upp (open circles) solutions. Solid line shows the $\eta=1$
 value fitting the experimental data.}
    \end{center}
                \label{etaall}
  \end{figure}

Obviously we cannot obtain constant fits to $\eta$ of a similar quality
for the "steep" solutions as those fits to the "flat" solutions. In order to
probe rather strong energy variation of $\eta$ in the former solutions we have
tried to fit it by polynomials of different powers. We treat these fits only 
as an ad hoc description of data in the particular region between 600 and
1000 MeV. It turned out that in order
to obtain good fits we need polynomials of the order as high as four (see 
Fig. 1a). Once again one can see the similar 
dependence of inelasticities when one compares 
a shape of the fitted curves to the "up-steep" and "down-steep" data points.
A minimum of $\eta$ for $m_{\pi\pi}$ around 800 MeV, indicating an
inelasticity, will be discussed below.  
Before passing to a physical discussion of possible consequences
of the strong energy dependence of inelasticity for the
"up-steep" solution below the $K\bar K$ threshold, let us discuss
a possibility of the fluctuation of  $\eta$ values around unity.
It is true that the deviation is not significant (see paper I:
$\chi^2=15$ for 17 points below 940 MeV) {\it if we ignore the shape}
of the $\eta$ distribution. However, the chance of all five
$\eta < 1$ points falling accidentally in the effective mass
region between 720 and 820 MeV is only $2\cdot 10^{-3}$. This
value was calculated by taking into account a
number of all the possibilities to choose five  lowest values of $\eta$
among twenty available points in the effective mass range between
600 and 1000 MeV and then grouping  them together in the range $b$.

In paper I in addition to inelasticities we have calculated the
phase shift values. For both "flat"
solutions phase shifts grow slowly with $m_{\pi\pi}$.
For the "steep" solutions we have obtained rather fast increase
of the S-wave I=0 $\pi\pi$ phase shifts near 770 MeV -- the value
close to the $\rho$-meson mass. 
We have tentatively fitted inelasticities and phase shifts of the
solution "up-steep" by a single
resonance. This was done for \mpp $<940$ MeV in order to avoid a possible 
influence of the \fo. We additionally allowed a simultaneous change of all
$\eta$ values by the same factor $R_\eta$ since in paper I they
 were fixed only by
minimizing the $\Sigma_{i}(\eta_{i}-1)^{2}$ value. In fact a combined 
fit ($\chi^{2}/NDF=26/30$) yields 
the reduction factor $R_\eta = 0.71\pm0.10$ and the following resonance
 parameters:
$m=(754\pm5)$ MeV, $\Gamma=(162\pm9)$ MeV and 
$x=0.74^{+0.10}_{-0.08}$, 
$x$ being an elasticity of the resonance. The mass and the width agree 
with $m=(753\pm19)$ MeV, $\Gamma=(108\pm53)$ MeV claimed by Svec \cite{svec97},
on the basis of the same data. However, such a considerable inelasticity is 
inconsistent 
with the available experimental data on the $4\pi$ system which is the only 
kinematically possible channel.\\
\hspace*{6mm}The lowest mass of the $4\pi$ system is probably available in 
the central production due to a $1/m_{4\pi}^{2}$ flux factor. In fact the 
WA91 \cite{WA91} and WA102 \cite{WA102} collaborations have found  a tiny peak 
around $800~$ MeV in the mass distribution of their $4\pi$ system 
produced at $450$ GeV in the reaction 
\begin{equation}
pp\rightarrow p_{f}\pi^{+}\pi^{+}\pi^{-}\pi^{-}p_{s},
\label{reaction_pp4pi}
\end{equation}
where $p_{f}$ and $p_{s}$ stand for fast and slow proton, respectively. 
This is however well explained by the reflection from the 
$\eta^{'}\rightarrow \eta\pi^{+}\pi^{-}$ decay
with the loss of the slow $\pi^{0}$ from the subsequent 
$\eta\rightarrow\pi^{+}\pi^{-}\pi^{0}$ decay. The IHEP-IISN-LANL-LAPP
collaboration \cite{alde87} studying at low $|t|$ a reaction similar 
to reaction (1) i.e.
\begin{equation}
\pi^-p\rightarrow \pi^0\pi^0\pi^0\pi^0 n
\label{reaction_4pi0}
\end{equation}
has found that the effective mass distribution of their $4\pi^0$ system starts
only around $800~$ MeV and rises smoothly. The LRL group \cite{LRL} studying
the reaction
\begin{equation}
\pi^+p\rightarrow \pi^+\pi^+\pi^-\pi^-\Delta^{++}
\label{reaction_4pidelta}
\end{equation}
found hardly any events with a  $4\pi$ mass below $1~$ GeV.
The $4\pi$ mass spectrum from annihilation 
$\overline{p}N\rightarrow5\pi$ starts at even higher mass as shown in [18-21]
. Thus the non-zero inelasticity 
in the \ups  solution does not have any reasonable physical interpretation. In
next section we will assume $\eta \equiv 1$ .


\section{Another approach to the \pipi scalar-isoscalar phase shifts
\label{another}}

\bigskip
\hspace{0.6cm}
In Sect. 2 we have shown that the 4$\pi$
channel is very weak below 1000 MeV. Therefore now
we assume that the \pipi S-wave inelasticity is exactly
equal to unity up to 990 MeV and we shall make a new analysis of
the \pipi isoscalar-scalar phase shifts obtained from the
$\pi^-p \to \pi^+\pi^-n$ data at 17.2 GeV/c.
Let us recall that in paper I a separation of the S-wave pseudoscalar 
$A_0$ and pseudovector $B_0$ exchange amplitudes for the
production process was performed and that we have calculated the
S-wave $\pi^+\pi^- \to \pi^+\pi^-$ amplitude $a_S$ from the
following formula:
\be
a_S = K f A_0 .
\label{as}\ee
In (\ref{as}) K is the proportionality factor
\be
K = - {8\pi p_{\pi}\sqrt{sq_{\pi}}\over m_{\pi\pi}\sqrt{2\cdot{g^2\over 4\pi}}}
~{1\over 2M},\label{K}\ee
where $p_{\pi}$ is the incoming $\pi^-$ momentum in the $\pi^-p$
centre of mass frame, $s$ is the square of the total energy in the
same frame, $q_{\pi}$ is the final pion momentum in the \pipi
rest frame, ${g^2\over 4\pi}$ is the pion-nucleon coupling constant
(taken as 14.6 in paper I) and $M$ is the proton mass. The
coefficient $f$ is the complex correction factor which (when
averaged over the four momentum transfer squared) represents the
$t$-dependence of the pion-nucleon vertex function, the
off-shellness of the exchanged pion and a possible phase of the
pion-exchange propagator -- a case where at high energy the
exchanged pion is treated as a Regge particle. This coefficient
was calculated  from the requirement that the sum $\sum_{i}
(\eta_{i}-1)^2$ was minimal for the inelasticities of
the scalar-isoscalar \pipi amplitude $a_0$ for a set of points
depending on the \pipi effective mass up to the \KK threshold
mass. The amplitude $a_0$ is connected to the amplitude $a_S$
and the isospin 2 S-wave amplitude $a_2$ in the following way:
\be
a_0 = 3a_S - {1\over 2} a_2.\label{a0}
\ee
The $a_0$ is also related  to the isospin 0 S-wave phase shift
$\delta_0$ and inelasticity $\eta$:

\be
a_0= {\eta e^{2i\delta_0} - 1 \over 2i}.
                                             \label{a0et}\ee
Motivated by the results of Sect. 2
we impose now the condition $\eta \equiv 1$ in the whole \mpp
range below 1 GeV. Then, from (\ref{a0et}) 

\be
a_0 = \sin \delta_0 e^{i\delta_0},\label{a0sin}\ee
so the modulus of $a_0$ is uniquely related to the phase shift
value $\delta_0$. Since in (\ref{as})  $A_0$ is the complex amplitude
calculated with some errors coming from experimental
uncertainties, then (\ref{a0}) and (\ref{a0sin}) are not necessarily
satisfied if we require $\eta \equiv 1$. In order to keep this
assumption valid we have to include at least one additional real
factor $n$ in equation (\ref{as}) at each $m_{\pi\pi}$, so now

\be
a^{new}_S = n a_S.                                   \label{anew}
\ee

We shall also assume that the isospin 2 amplitude $a_2$ is fully
elastic and can be described by the corresponding phase shift
$\delta_2$ 

\be
a_2 = \sin \delta_2 e^{i\delta_2}.                        \label{a2}
\ee
Then, following (\ref{a0et}) and inserting (\ref{anew}) into (\ref{a0})
one has to satisfy equations

\be
\eta^2 = \left\vert 1+2ia_0\right\vert^2 =
\left\vert1+2i(3KfnA_0-{1\over 2}a_2)\right\vert^2 \equiv1.  \label{eta2}
\ee 
We treat (\ref{eta2}) as a quadratic equation for $n$. Its roots are

\be
n = {1\over 6\left\vert a_S\right\vert}(b \pm
\sqrt{b^2-3\sin^2\delta_2}),                                \label{en}
\ee 
where
\be
b = (1+\sin^2\delta_2) \sin\alpha + {1\over
2}\sin^2\delta_2\cos\alpha                                    \label{b}
\ee 
and $\alpha$ denotes the phase of $a_S$ :

\be
a_S = \left\vert a_S\right\vert e^{i\alpha}.\label{aSa}\ee 

Let us remark that in the limit of $\delta_2$ going to 0
(vanishing $a_2$) we should consider only the upper sign (+) in
(\ref{en}). In general, both solutions for $n$ are
possible; we have, however, used only the root with the upper sign 
since its value was closer to unity. The isotensor phase
shift $\delta_2$ is calculated according to the parametrization
given in paper I which fits well the data of Ref. \cite{hoogland} obtained by 
method B. 

For each value of $m_{\pi\pi}$ we have to check
whether the roots exist. With $\eta = 1$ 
the amplitude $a_0$ must satisfy the elastic unita\-ri\-ty condition 
\be
Im ~a_0 = \left\vert a_0\right\vert^2.    \label{unitar} 
\ee
Therefore the following inequality must be fulfilled

\be
\left\vert b\right\vert \ge \sqrt{3} \left\vert \sin \delta_2
\right\vert.\label{bw}\ee 

We have obtained the following numerical results for the
solutions discussed in I: the inequality (\ref{bw}) was
satisfied for all twenty \mpp points of the solution \upp and
for 19 points (except of the extreme point at 990 MeV)
corresponding to the solution \downp. However for 7 points of
the solution \ups and for 12 points of the solution \downs
the condition (\ref{bw}) was violated. This fact casts a serious doubt on
a validity of both "steep" solutions. The resulting values of $n$
calculated in cases when (\ref{bw}) was satisfied are shown in
Fig. 2. The errors of $n$ are due to experimental errors of the
modulus $\left\vert a_S\right\vert$ and the phase $\alpha$
extracted from experiment. No errors of $\delta_2$ were taken
into account since $\delta_2$ values were calculated using the
smooth theoretical parametrization. We see in Fig. 2b
that both "flat" solutions are well fitted by constants
very close to unity ($0.994 \pm 0.03$ for the \downp
and $0.997 \pm 0.04$ for the \upp solution). On the other hand a
variation of $n$ with \mpp for two "steep" solutions is better
described by a parabola than by a constant (see Fig. 2a). Such
strong dependence on \mpp of the coefficient $n$ corresponding
to both "steep" solutions can be used as a fairly strong
argument against an acceptance of these solutions as good
physical solutions.

\begin{figure}
    \begin{center}
\xslide{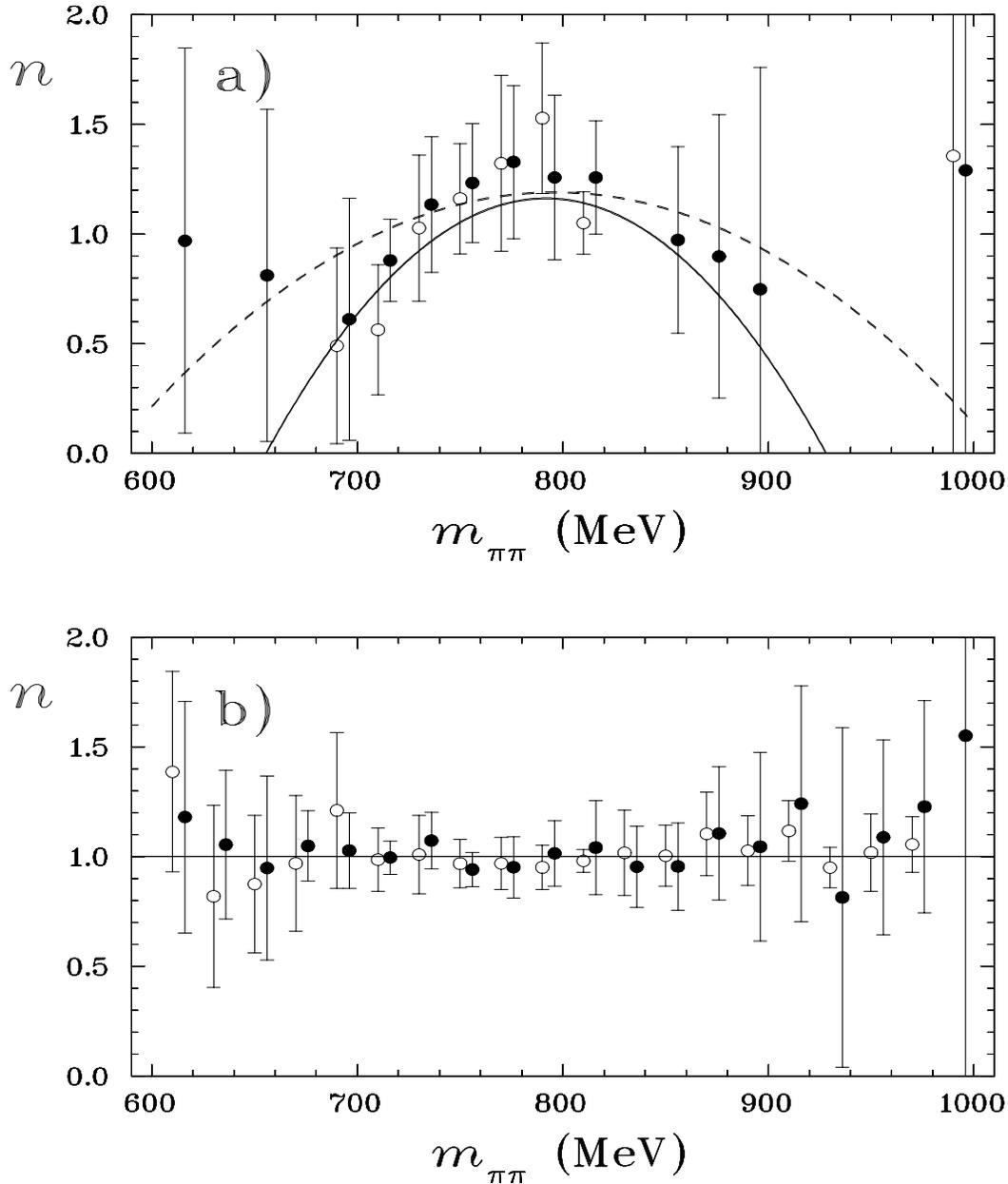}{16.5cm}{55}{52}{555}{723}{14cm}

\vspace{0.5cm}

\caption{{\bf a)} Effective mass dependence of the parameter $n$ for the 
\downs (full circles) and \ups (open circles, shifted by 6 MeV) solutions. 
Solid and dashed lines
 represent fits of the second order polynomial to $n$ for the \downs
and \ups solutions, respectively. Note that in many bins no physical solution 
could be found. {\bf b)} Same as in a) but for the \downp (full circles) 
and \upp (open circles) solutions. Solid line represents
values of the constant parameters fitted to $n$ for the \downp and \upp 
 solutions.}
    \end{center}
                \label{nall}
  \end{figure}

For completeness we present in Fig. 3 new \pipi phase shifts
calculated from (\ref{a0sin}) using $a^{new}_S$ given by
(\ref{anew}). Obviously we show only these points for
which the corresponding values of $n$ do exist. The new phase
shifts, calculated under the assumption that $\eta \equiv 1$,
agree very well with those presented in paper I for the "flat" solutions.
  For the "steep" solutions this agreement is not so good
and the new errors of $\delta_0$ are larger than those shown
in paper I.

\begin{figure}
    \begin{center}
\xslide{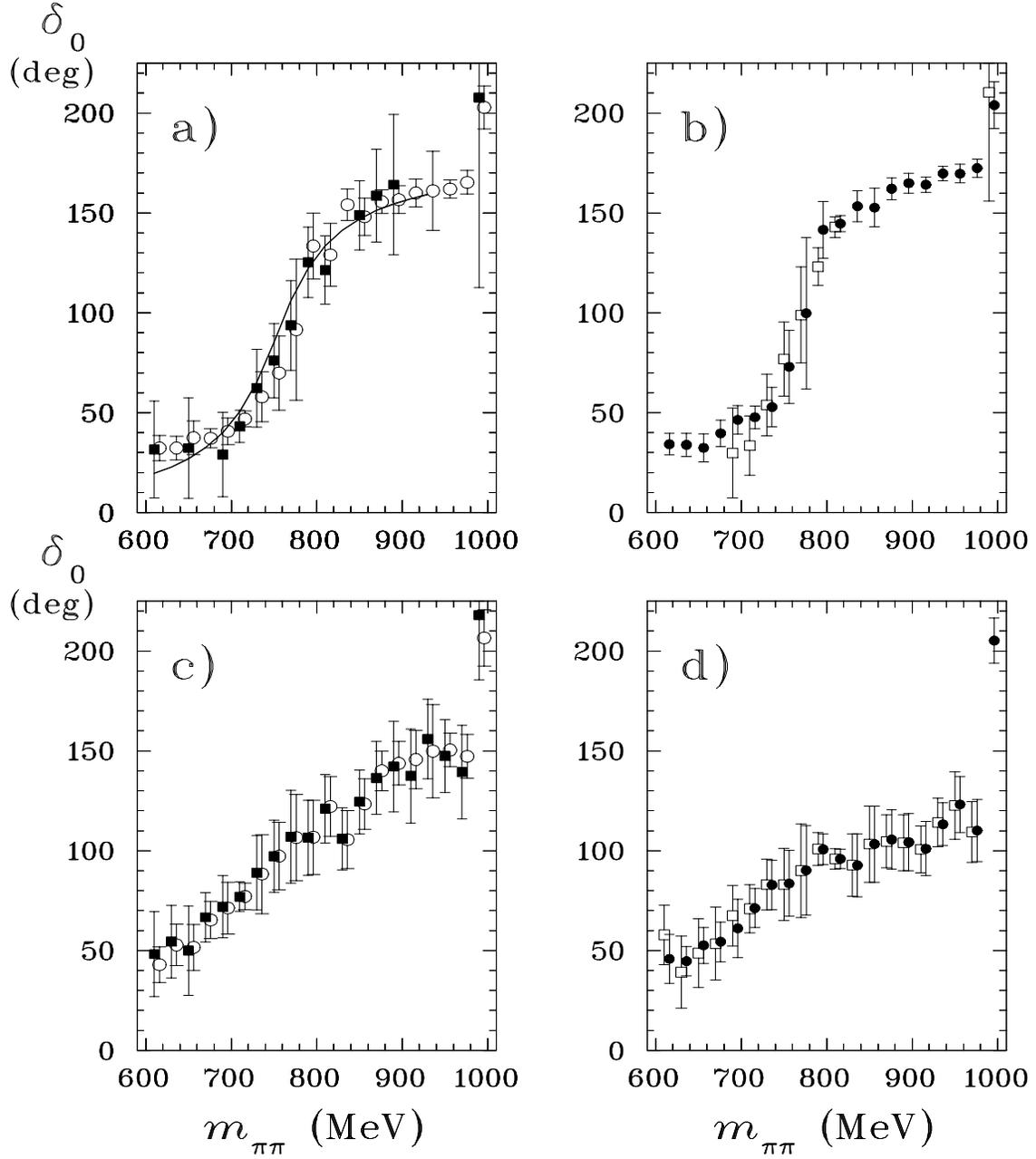}{17cm}{35}{74}{575}{745}{15cm}
\caption{Comparison of the scalar--isoscalar \pipi phase shifts
obtained in paper I (circles, shifted in \mpp by 6 MeV) with the 
new phase shifts (squares) calculated under the 
assumption $\eta \equiv 1$. {\bf a)} For the \ups solution. 
Solid line shows the Breit-Wigner fit to the data marked by open circles
as described in Sect. 2. {\bf b)} For the \downs 
solution. {\bf c)} For the \upp solution.
\mbox{{\bf d)} For the \downp solution.}}
    \end{center}
                \label{abcd}
  \end{figure}


\section{Discussion \label{summary}}

\bigskip
\hspace{0.6cm}
In Sections 2 and 3 we have presented arguments that both "steep"
solutions have unphysical behaviour. On the other hand both "flat"
solutions satisfy well our tests and none of them can be eliminated using 
the methods described in this paper. Therefore
let us discuss common features of these solutions and  
 major differences between them. In Fig. 4 the "flat" solutions are plotted  
in a wide effective mass range up to 1600 MeV. Their shape is quite similar.
One sees an initial steady grow of phase shifts with \mpp above 600 MeV, then
 at about
1000 MeV, corresponding to the \KK threshold, there is a jump as high as 
$140^0$ and further on a fairly steep increase above 1300 MeV. An interpretation
of this behaviour of phases in terms of three scalar resonances $ f_0(500)$, \fo
and \fgg, started in paper I, has been continued in more detail in Refs. 
[23,24]
. We do not repeat it here but we underline
major differences between the "up-flat" and "down-flat" phase shifts since they
lead to different values of the $ f_0(500)$ resonance parameters. The 
$ f_0(500)$ mass for the "up-flat" solution is by about 50 MeV higher than
the corresponding mass for the "down-flat" solution. The reversed relation
for the $ f_0(500)$ width leading to a difference between 45 and 50 MeV is also 
observed. We do not see
important differences between the \fo parameters for the above solutions.
  As seen
in Fig. 4 the most important differences, reaching about $45^{0}$,
exist between 800 and 1000 MeV. They are related to a difference between
the {\em moduli} of the S-wave pseudoscalar amplitude (see paper I). This
difference of moduli is then directly transformed into a difference between 
the phase
shifts because inelasticity for both "flat" solutions below the \KK threshold
is very close to unity. This fact in turn guarantees a fulfilment of the
elastic unitarity condition (\ref{unitar}). Closer inspection into Fig. 4
allows one to see a continuity of the phase differences above the \KK
threshold, namely the "up-flat"
points lie systematically above the "down-flat" points up to about 1300 MeV.
Above this value of \mpp we do not observe any systematic difference. 

\begin{figure}
    \begin{center}
\xslide{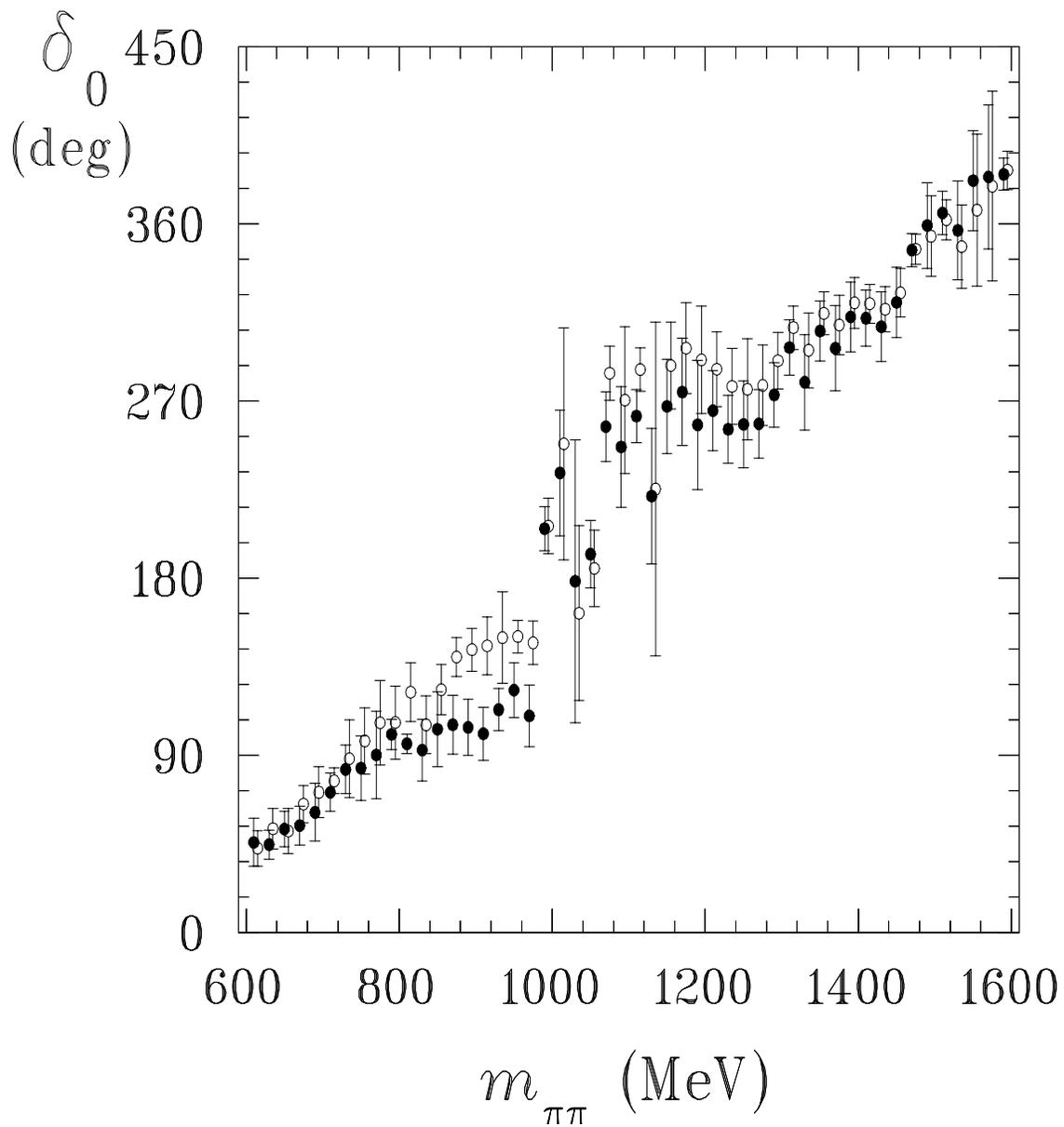}{16cm}{48}{142}{564}{612}{15cm}

\vspace{1cm}

\caption{Comparison of the scalar--isoscalar \pipi phase shifts 
obtained in \mbox{paper I} for the \downp (full circles) and \upp (open circles) 
solutions.}
    \end{center}
                \label{delta}
  \end{figure}

Since the data points of the \upp solution lie between the points of the
(already excluded) \ups solution and the \downp one, we have checked whether
 the 
$f_0(750)$ survives in this solution. This was done by fitting elasticities and
 phase shifts of the \upp solution by a single Breit-Wigner term like it was 
 done in Sect. 2 for the \ups solution. No overall change of $\eta$ values was 
 needed ($R_\eta=1.01^{+0.07}_{-0.15}$) and the resonance parameters are 
 $m=(732\pm8)$ MeV, $\Gamma=(246^{+37}_{-25})$ MeV and $x=1.00^{+0.08}_{-0.16}$.
The large width is inconsistent with a narrow $f_0(750)$.


\section{Summary}
\hspace{0.6cm}
In conclusion, we have studied in detail the \pipi effective mass dependence of 
the S-wave isoscalar phase shifts corresponding to four solutions "up-steep",
"down-steep", "up-flat" and "down-flat" found in paper I. Both "steep" solutions
exhibit an inelasticity behaviour which has no physical interpretation. We do
not find any data on the $4\pi$ systems which could explain a  strong \mpp
dependence of the inelasticity corresponding to the "steep" solutions
below the \KK threshold. The "down-steep" solution was already rejected in
paper I since its inelasticity substantially exceeded unity for \mpp above 820
 MeV.
Assuming that the "up-steep" inelasticity is an unusual fluctuation we impose
$\eta \equiv 1$ for all points and, for completeness, in all solutions. This leads to non-physical 
results in 7/20 mass bins for the "up-steep" solution and in 12/20 bins
for the
"down-steep" solution. In the remaining mass bins the parameters behave in a 
queer way (compare Figs. 2a and 2b ). We conclude that the "up-steep"
solution cannot be treated as a good physical set of phase shifts. It can be
eliminated together with the "down-steep" solution. However, the "up-flat" and "down-flat" 
solutions easily
pass our tests. We would like to stress that both the $f_0(500)$ and \fo 
resonances are present in the "flat" solutions. This is not true for a 
relatively narrow \ep.


\addvspace{1cm}
{\em Acknowledgements.} The authors are very grateful to Dr. Michael 
Pennington whose suggestion initiated this study. R. Kami\'nski thanks
NATO for the grant in 1999.


\end{document}